\documentclass[%
 aps,
 prapplied,
 amsmath,amssymb,
superscriptaddress,
 reprint,%
]{revtex4-2}

\usepackage{ mathrsfs  }
\usepackage{graphicx}
\usepackage{dcolumn}
\usepackage{bm}
\usepackage[dvipsnames]{xcolor}

\usepackage{gensymb}

\newcommand{\dd}{\mathrm{d}}
\newcommand{\gp}{\dot\gamma}

\newcommand{\sm}{\sigma_{\text{M}}}

\newcommand{\sy}{\sigma_\text{y}}
\newcommand{\Gp}{\dot\Gamma}
\newcommand{\tit}{\tilde{t}}
\newcommand{\tm}{t_\text{M}}
\newcommand{\titm}{\tit_\text{M}}

\newcommand{\Sm}{\Sigma_\text{M}}
\newcommand{\Sy}{1}
\newcommand{\Tf}{T_\text{f}}
\newcommand{\tf}{\tau_\text{f}}
\newcommand{\lb}{\ell_b}

\usepackage{color}
\usepackage{tikz}
\usetikzlibrary{shapes}

\newcommand{\circlelightgoldenrod}{\raisebox{0.5pt}{\tikz{\node[draw,scale=0.5,circle,fill=white!60!yellow](){};}}}

\newcommand{\triangledarkgoldenrod}{\raisebox{0.5pt}{\tikz{\node[draw,scale=0.3,regular polygon, regular polygon sides=3,fill=black!0!brown,rotate=0](){};}}}

\newcommand{\trianglelightgoldenrod}{\raisebox{0.5pt}{\tikz{\node[draw,scale=0.3,regular polygon, regular polygon sides=3,fill=white!60!yellow,rotate=0](){};}}}

\newcommand{\diamondsiennafour}{\raisebox{0.5pt}{\tikz{\node[draw,scale=0.4,diamond,fill=black!40!brown](){};}}}

\newcommand{\squaredarkgoldenrod}{\raisebox{0.5pt}{\tikz{\node[draw,scale=0.4,regular polygon, regular polygon sides=4,fill=black!0!brown](){};}}}

\newcommand{\invertedtriangledarkgoldenrod}{\raisebox{0pt}{\tikz{\node[draw,scale=0.3,regular polygon, regular polygon sides=3,fill=black!0!brown,rotate=180](){};}}}

\definecolor{fig1_1}{rgb}{1.000000000000000 ,  0.781200000000000  , 0.497500000000000}
\definecolor{green(ncs)}{rgb}{0.0, 0.62, 0.42}
\definecolor{deepskyblue}{rgb}{0.0, 0.75, 1.0}

\begin{document}

\title{Continuum modeling of Soft Glassy Materials under shear}

\author{Roberto Benzi}
\affiliation{Dipartimento di Fisica and INFN, Università di Roma “Tor Vergata”, Via della Ricerca Scientifica, 1-00133 Rome, Italy\looseness=-1}
\author{Thibaut Divoux}
\affiliation{ENSL, CNRS, Laboratoire de physique, F-69342 Lyon, France\looseness=-1}
\author{Catherine Barentin}
\affiliation{Univ.~de Lyon, Université Claude Bernard Lyon 1, CNRS, Institut Lumière Matière, F-69622 Villeurbanne, France\looseness=-1}
\author{S\'ebastien Manneville}
\affiliation{ENSL, CNRS, Laboratoire de physique, F-69342 Lyon, France\looseness=-1}
\affiliation{Institut Universitaire de France (IUF)}
\author{Mauro Sbragaglia}
\affiliation{Dipartimento di Fisica and INFN, Università di Roma “Tor Vergata”, Via della Ricerca Scientifica, 1-00133 Rome, Italy\looseness=-1}
\author{Federico Toschi}
\affiliation{Department of Applied Physics, Eindhoven University of Technology, P.O. Box 513, 9 5600 MB Eindhoven, Netherlands and CNR-IAC, Rome, Italy\looseness=-1}

\begin{abstract}
Soft Glassy Materials (SGM) consist in dense amorphous assemblies of colloidal particles of multiple shapes, elasticity, and interactions, which confer upon them solid-like properties at rest. They are ubiquitously encountered in modern engineering, including additive manufacturing,  semi-solid flow cells, dip-coating, adhesive locomotion, where they are subjected to complex mechanical histories. Such processes often include a solid-to-liquid transition induced by large enough shear, which results in complex transient phenomena such as non-monotonic stress responses, i.e., stress overshoot, and spatially heterogeneous flows, e.g., shear-banding or brittle failure. In the present article, we propose a pedagogical introduction to a continuum model based on a spatially-resolved fluidity approach that we recently introduced to rationalize shear-induced
yielding in SGMs. Our model, which relies upon non-local effects, quantitatively captures salient features associated with such complex flows, including the rate dependence of the stress overshoot, as well as transient shear-banded flows together
with nontrivial scaling laws for fluidization times. This approach offers a versatile framework to account for subtle effects, such as avalanche-like phenomena, or the impact of boundary conditions, which we illustrate by including in our model the elasto-hydrodynamic slippage of soft particles compressed against solid surfaces.
\end{abstract}

\maketitle

\section{Introduction}

Soft Glassy Materials (SGMs) encompass a broad variety of colloidal particles densely packed into an amorphous microstructure showing solid-like properties at rest~\cite{Bonnecaze:2010,Joshi:2014}. These particles, which can be either soft and deformable or hard, form a jammed assembly with glassy-like mechanical properties characterized by ($i$)~a linear viscoelastic response where the elastic contribution is dominant~\cite{Sollich:1997}, and ($ii$)~time-dependent properties referred to as ``aging" in the literature~\cite{Cipelletti:2000,Joshi:2018}. Moreover, under a sufficiently large external stress or strain, particles can rearrange. For vanishingly low shear rates, such rearrangements take the form of local plastic events such as T1 events in foams and emulsions~\cite{Princen:1986}, or shear-transformation zones in colloids~\cite{Schall:2007}. Eventually, for sufficiently large accumulated deformation, these plastic events, which act as a mechanical noise, lead to the fluidization of the sample. Remarkably, such a shear-induced solid-to-liquid transition displays generic features that are quite insensitive to the sample microstructure~\cite{Rodney:2011,Falk:2011,Bonn:2017,Nicolas:2018}. For instance, under a constant applied shear rate $\dot \gamma$, the stress $\sigma$ builds up and reaches a maximum before relaxing towards a steady-state value. Such a non-monotonic response, known as a \textit{stress overshoot}~\cite{Mewis:2009}, coincides with the yielding of the sample, which may either flow homogeneously, or rather display a spatially heterogeneous yielding process~\cite{Divoux:2016}. In the latter case, flow heterogeneity occurs due to localized, brittle-like failure~\cite{Magnin:1990}, or results from a more ductile process in which an arrested region coexists with a fluidized one, referred to as a ``shear band,'' whose lifespan depends on the volume fraction and on the particle interactions~\cite{Gopalakrishnan:2007,Besseling:2010,Divoux:2011b}. 
\begin{figure*}[t!]
  \centering
  \includegraphics[width=18cm]{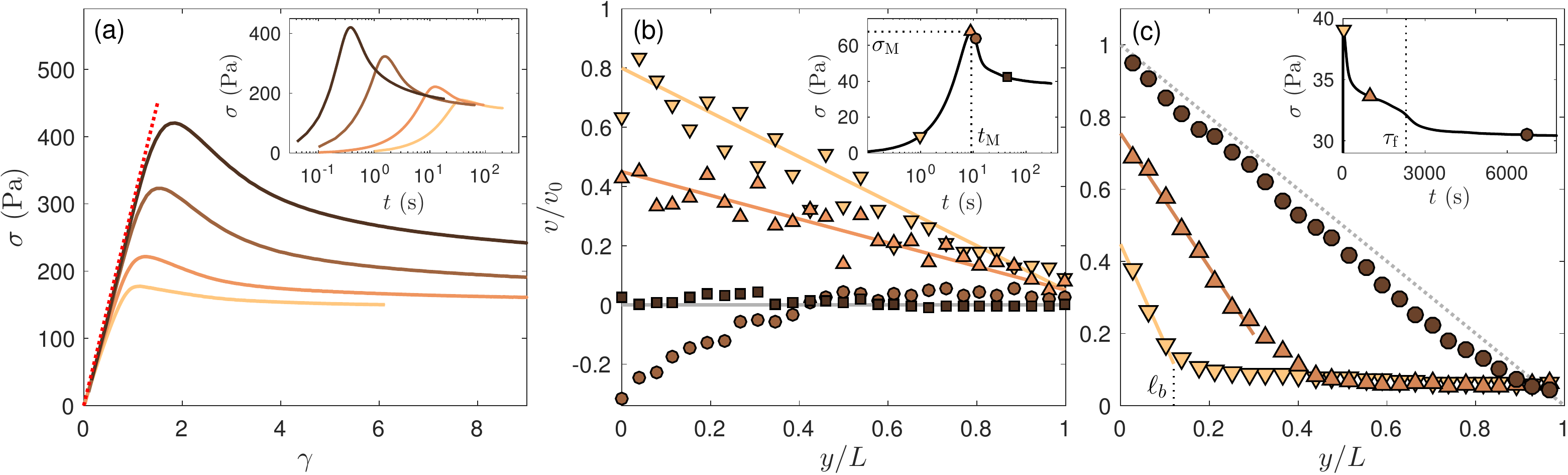}
  \caption{{\bf Phenomenology of shear start-up experiments in Carbopol microgels.} (a)~Stress $\sigma$ as a function of strain $\gamma=\gp t$ recorded after a shear rate $\gp = 5$, 1, 0.2, and 0.03~s$^{-1}$ from top (darker color) to bottom (lighter color) is applied at time $t=0$. The red dashed line highlights the linear response at short time $\sigma=G_0\gamma$ with $G_0=300$~Pa. Inset: same data plotted as a function of time $t$ using semilogarithmic scales.
  (b,c)~Velocity profiles $v$ normalized by the velocity of the moving plate $v_0$ as a function of the distance $y$ to the moving wall normalized by the gap size $L$ and recorded (b)~at short times around the stress overshoot under $\gp=0.1$~s$^{-1}$, and (c)~at long times during the transient shear-banding regime under $\gp=0.7$~s$^{-1}$. In both cases, the inset shows the corresponding stress response $\sigma(t)$ and the colored symbols show the times at which the velocity profiles in the main graph are recorded. The colored lines are guides to the eye in (b) and fits to the velocity profile in the shear band in (c). The gray dashed line in (c) shows the velocity profile expected for a Newtonian fluid in the absence of wall slip.
  }
  \label{fig_experiments}
\end{figure*}

Various modelling efforts have been undertaken over a broad range of spatial scales, from that of the building block, thanks to, e.g., Molecular Dynamic simulations~\cite{Ozawa:2018,Vasisht:2020,Vasisht:2020b}, to mesoscale or macroscopic continuum approaches in which the SGM microstructure is accounted for only by a few parameters~\cite{Mewis:2009,SouzaMendes:2011}, up to typically 10, in order to capture more subtle effects such as non-isotropic resistance of the sample inherited from shear history~\cite{Dimitriou:2014,Geri:2017}. Here, we shall focus on a continuum approach 
traditionally 
referred to as ``fluidity models'' \cite{Picard:2002,Derec:2003} in which the microscopic properties of the sample are expressed by the fluidity $f$, a local quantity, which stands for a rate of plastic events. 

Recently, fluidity models were derived theoretically by Bocquet {\it et al.} \cite{Bocquet:2009} as the continuum limit of a microscopic equation for the probability distribution originally proposed by H\'ebraud and Lequeux \cite{Hebraud:1998}. Such an approach showed that fluidity models naturally encompass non-local effects in steady state via a so-called ``cooperativity'' length scale that quantifies the extension of the region that is impacted by a neighboring plastic rearrangement~\cite{Goyon:2008,Mansard:2012,Geraud:2013,Geraud:2017}. The approach of Ref.~\cite{Bocquet:2009} was extended to transient flows by some of us in Ref.~\cite{Benzi:2016}. This extended non-local version of the fluidity model has been used to quantitatively capture the key features of the yielding transition of a soft glass~\cite{Benzi:2019,Benzi:2021PRL,Benzi:2021PRE}.

In the present Perspective, we first summarize these findings to illustrate the power of the non-local fluidity model. Second, we extend our approach and include the elasto-hydrodynamic slippage of soft particles compressed against a solid surface in order to describe recent results from the literature~\cite{Khabaz:2021}.


\section{{Phenomenology of shear start-up experiments}}

In a shear start-up experiment, one imposes a constant shear rate $\dot{\gamma}$ at time $t=0$ upon an SGM  initially at rest. Figure~\ref{fig_experiments} illustrates the general phenomenology of shear start-up through a selection of experimental results on Carbopol microgels~\cite{Divoux:2010,Divoux:2011,Divoux:2012}. As recalled in the introduction, a {\it stress overshoot} is classically observed: at short times $t$, the shear stress $\sigma$ grows linearly with the strain $\gamma=\dot \gamma t$ [see red dashed line in Fig.~\ref{fig_experiments}(a)], which is typical of an elastic response. At longer times, the stress progressively deviates from a linear response and reaches a maximum $\sm$ at time $\tm$. As the stress maximum is reached, the material has become strongly anisotropic, and subsequent stress relaxation processes lead the material to flow on even longer time scales. This global behaviour is typical of ductile-like yielding. 

To get more insight into the local structure of the flow during the solid-to-liquid transition, velocity profiles measured using ultrasonic velocimetry are displayed in Fig.~\ref{fig_experiments}(b) and \ref{fig_experiments}(c). In the case of the present microgels, the material is homogeneously strained prior to the stress overshoot, and the stress maximum corresponds to the point when the microgel fails at the shearing surface [see Fig.~\ref{fig_experiments}(b)]. Such failure is followed by an elastic recoil [see the negative velocities for the velocity profile with $\circ$ symbols in Fig.~\ref{fig_experiments}(b)], then by a fully arrested regime with $v=0$ across the whole sample except for a thin, unresolved lubrication layer at the moving wall. Yet, on time scales much longer than the time $\tm$ of the stress maximum, a fluidized region, i.e., a shear band, of width $\lb$ grows from the moving wall and coexists with the arrested, solid-like material until a fully homogeneous flowing state is reached at a well-defined \textit{fluidization time} $\tf$ [see Fig.~\ref{fig_experiments}(c)]. 
Finally, whatever the complexity of the stress relaxation, the stationary velocity profiles of the present microgels are ultimately homogeneous with insignificant wall slip. It is also essential to note that the fully flowing material is well described by the widespread Herschel-Bulkley (HB) constitutive law relating the stress $\sigma$ and the shear rate $\dot \gamma$ in the system at steady state~\cite{Divoux:2011b},
$\sigma = \sy + A\, {\dot \gamma}^{n}$,
where $\sy$ is the yield stress, $A$ the consistency and $n$ the shear-thinning index. 

From the experimental results displayed in Fig.~\ref{fig_experiments}, one may argue that the yielding transition can be considered as a dynamical first-order phase transition, where one phase (the fluid-like phase) nucleates into the other phase (the solid-like phase). As we shall see, this idea underlies most of the following discussion, which focuses on two relatively simple yet fundamental questions: ($i$) How does the stress overshoot $\sm$ depend on the applied shear rate $\dot \gamma$? ($ii$) How does the fluidization time $\tf$ depend on the applied shear rate $\dot \gamma$ or stress $\sigma$?
Our ultimate goal is to  obtain a general framework that describes  quantitatively the yielding transition and its mesoscopic features. 

\section{Continuum modeling}
We start by considering a two-dimensional shear geometry where the SGM is confined between two infinite parallel plates separated by a distance $L$. The flow is assumed to be one-dimensional along the direction $x$, i.e., it is described by a velocity field $\mathbf{v}=(v_x,v_y)$ with $v_x(x,y,t)=v(y,t)$ and $v_y(x,y,t)=0$, where $y$ denotes the velocity gradient direction and $t$ the time. The wall at $y=0$ moves with a constant velocity $v_0$ imposed at the initial time $t=0$, while the wall at $y=L$ remains fixed with zero velocity. As in experiments, the SGM is assumed to be initially at rest so that $v(y,0)=0$ for $y\in[0,L]$. For the sake of simplicity, we introduce the dimensionless stress and shear rate, $\Sigma=\sigma/\sy$ and $\Gp=\gp/(\sy/A)^{1/n}$, such that the SGM in steady state follows the dimensionless HB law: 
\begin{equation}\label{eq:simplifiedHB}
\Sigma(\Gp)=1+\Gp^{n}\,.
\end{equation}
Following Refs.~\cite{Goyon:2008,Bocquet:2009}, in order to describe the local behaviour of the SGM, we introduce the fluidity $f(y,t)$ of the SGM as the relevant order parameter in the system, as well as a characteristic length scale, called the {\it cooperativity scale} $\xi$, which controls spatial dynamics of the fluidity. Qualitatively, the fluidity corresponds to the rate of plastic events at a given time and position in the system. When the SGM flows in steady-state under an applied stress $\Sigma$, Bocquet {\it et al.}~\cite{Bocquet:2009} linked  the fluidity to elasto-plasticity at mesoscale through the following equation:
\begin{equation} \label{eq:Goyon}
\xi^2 \Delta f + m f - f^{3/2} = 0\,,
\end{equation}
where 
\begin{equation}\label{eq:m}
m^2 \equiv \frac{( \Sigma- \Sy)^{1/n}} {\Sigma} \,\Theta(\Sigma-1)\,,
\end{equation}
and $\Theta$ is the Heaviside function. As noted in Ref.~\cite{Bocquet:2009}, Eq.~\eqref{eq:Goyon} can be associated to  the functional derivative of 
\begin{equation}
\label{5} F[f] =  \int_0^L \Phi[f]\, dy   \equiv \int_0^L \left[ \frac{1}{2} (\nabla f)^2 - \frac{1}{2} m f^2 + \frac{2}{5} f^{5/2} \right]\,\dd y.
\end{equation}

\begin{figure*}[t!]
  \centering
  \includegraphics[width=0.67\columnwidth]{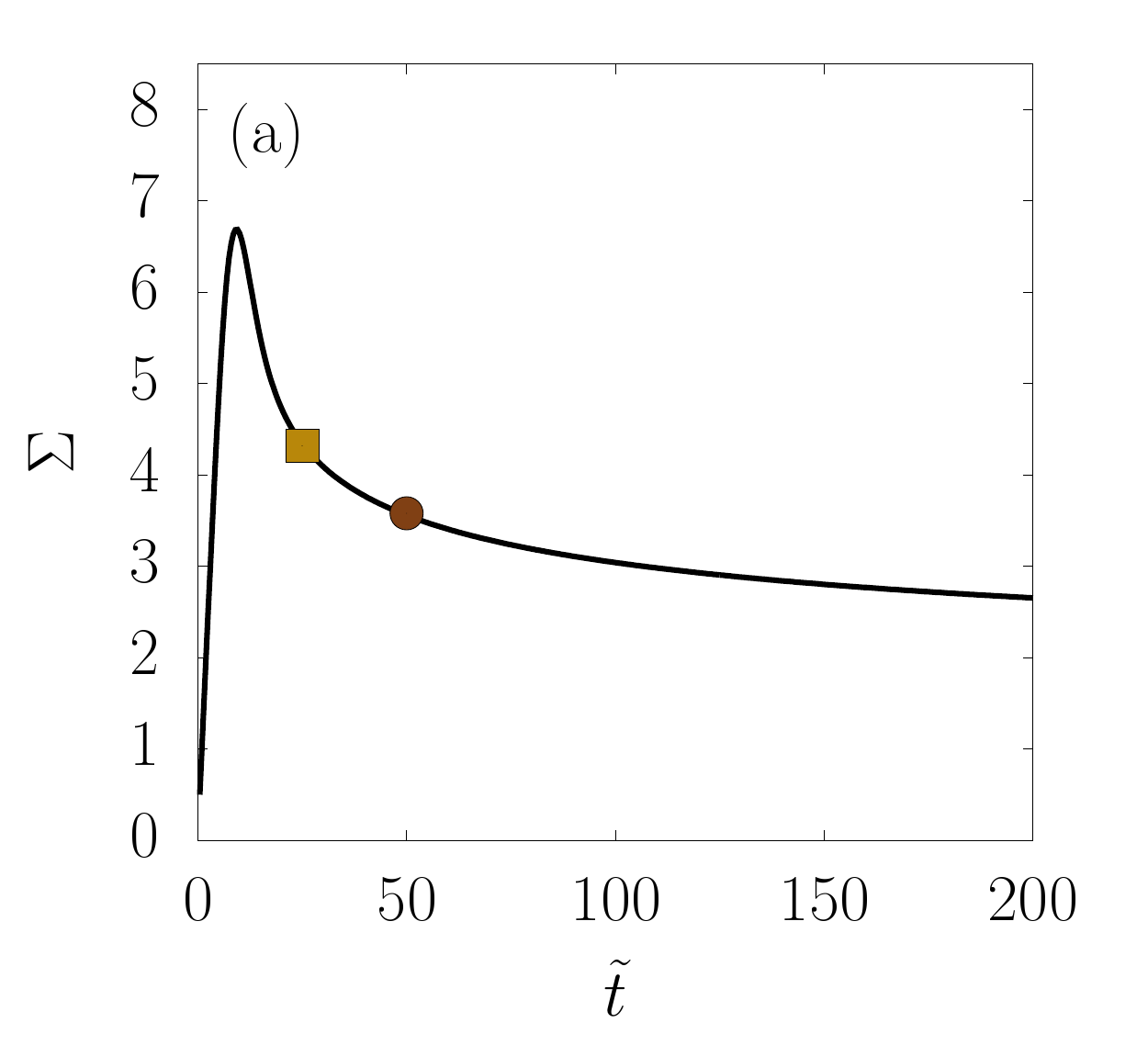}
  \includegraphics[width=0.67\columnwidth]{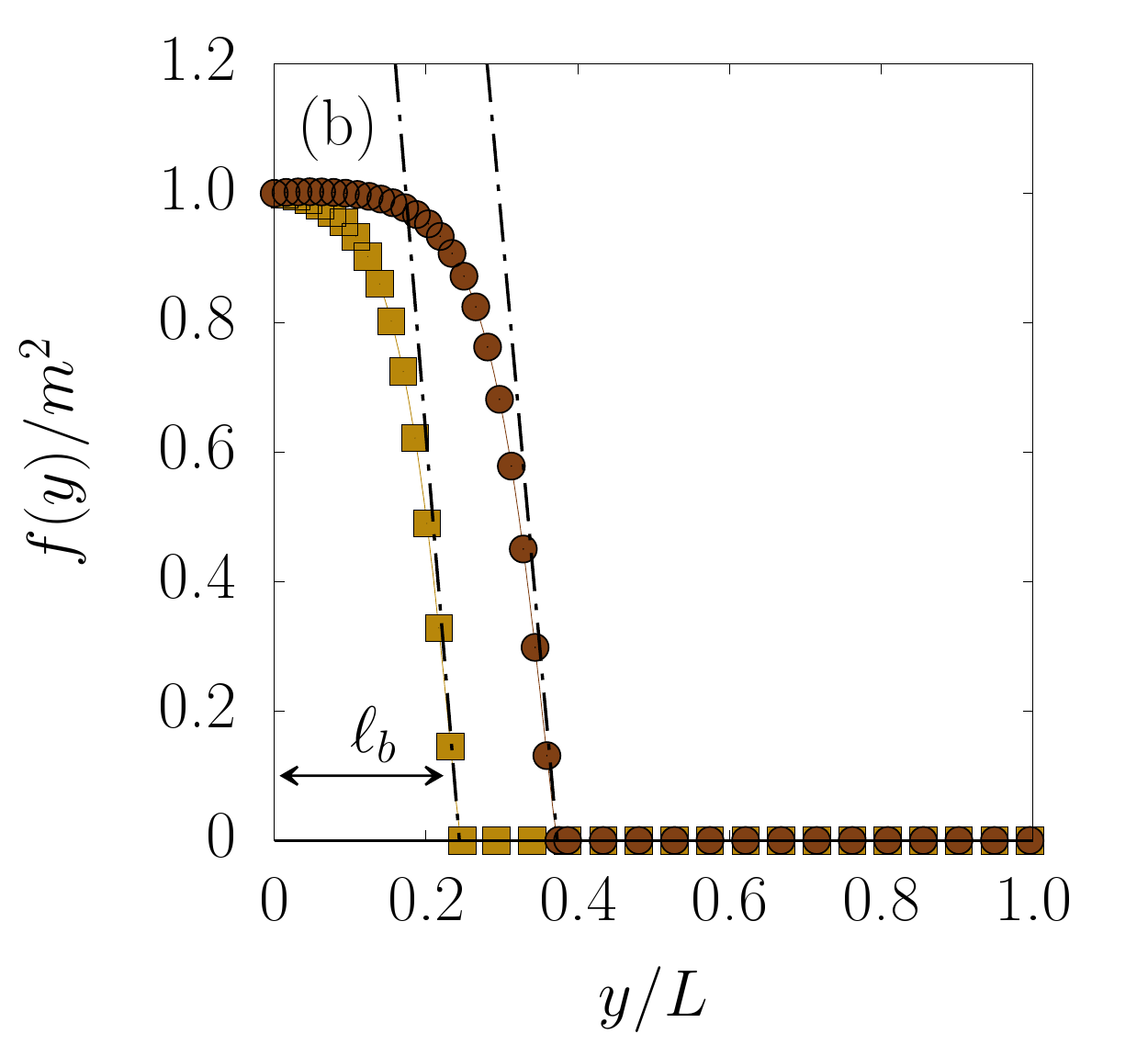}
  \includegraphics[width=0.67\columnwidth]{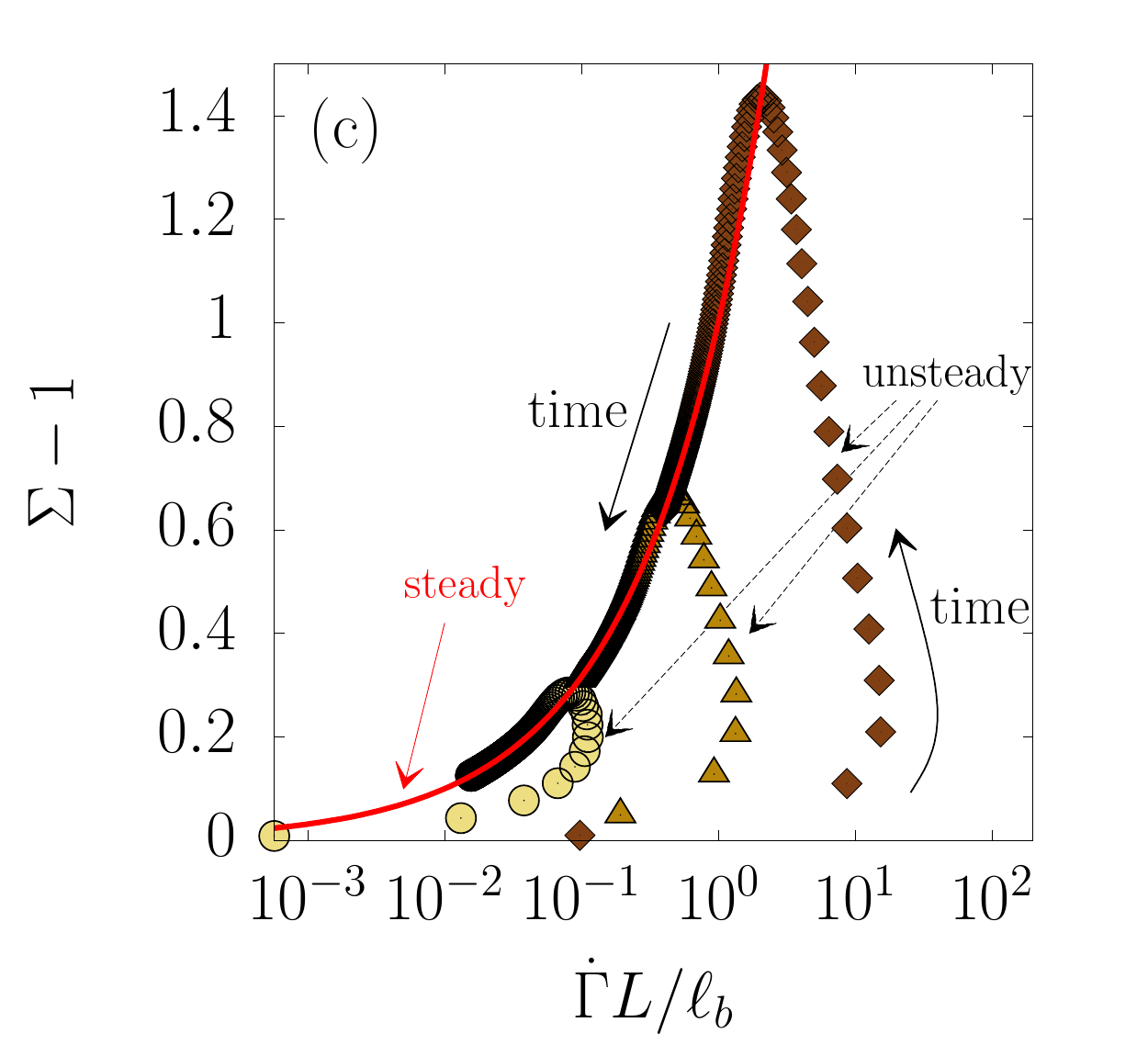}
    \caption{{\bf Phenomenology of shear start-up in the fluidity model.} (a)~Stress response $\Sigma(\tilde{t})$ computed for $n=1/2$, $L=1$, $\tau=10$, $\xi = 0.04$, $\dot \Gamma=2$, and $f_0 = 10^{-4}$. The colored symbols highlight the times at which the data in (b) are extracted. (b)~Normalized fluidity profiles $f(y)/m^2$ as a function of the normalized spatial coordinate $y/L$. The dashed-dotted lines represent an analytical estimate of the steepness of the interface between the fluidized and the solid-like regions. (c)~Stress distance to the yield stress, $\Sigma-1$, as a function of the effective shear rate, $\dot\Gamma L/\lb$, developing in the shear band of width $\lb$ [see arrow in (b)], for various global shear rates $\Gp=8\times 10^{-4}$ (\protect \circlelightgoldenrod), $\Gp=9\times 10^{-3}$ (\protect \triangledarkgoldenrod) and $\Gp=10^{-2}$ (\protect \diamondsiennafour). The red solid line corresponds to the HB prediction, $\Sigma -1 = (\Gp L/\lb)^{1/2}$, relating the shear stress to the effective shear rate. 
    }
  \label{fig:band_evolution}
\end{figure*}

From the above equation, it is tempting to consider $F [f]$ as a {\it free energy} functional for the fluidity $f$. Based on this idea, we proposed in Refs.~\cite{Benzi:2016,Benzi:2019} to extend the approach and formulate the {\it dynamics} of  the system using $F[f]$. In order to model shear start-up, i.e., a constant velocity $v_0$ imposed at the moving wall at $t=0$, we take the shear rate $\dot \Gamma$ as the imposed control parameter. In this case, the quantity $ \tilde f \equiv f / \dot \Gamma$ should be proportional to the number of plastic events occurring at some position $y$ over the time scale $\dot \Gamma ^{-1}$. Such a number may increase or decrease locally depending on the dynamics of the system induced by the external driving $\dot \Gamma $. Since $\dot\Gamma$ is constant, the temporal variation $\partial \tilde f / \partial t$ is nothing but the fluidity variation due to $ \tilde t \equiv \dot \Gamma t $, i.e., $\partial \tilde f / \partial t = \partial f / \partial \tilde t$.  Our first modeling step is to assume that the fluidity dynamics is given by: 
\begin{equation}\label{eq:fluidity_mobility}
\frac{\partial f}{\partial \tilde t} = - \kappa[f] \frac{\delta F[f]}{\delta f}
\end{equation}
where $\kappa[f]$ plays the role of a ``mobility". 
Suppose now that the system can be decomposed into two different regions: a fluidized region where $f>0$ and a solid-like region where $f=0$. Our second important assumption is to require that both regions correspond to {\it stationary solutions} of Eq.~\eqref{eq:fluidity_mobility}. 
Assuming $\kappa[f]$ to be an analytic function of $f$, the simplest choice is $\kappa[f] \sim f$. This implies that the formation of a shear band in the system coexisting with a solid-like region with {\it exactly} $f=0$ can be described by the superposition of two stationary states of the dynamics: one corresponding to $f>0$ [see Eq.~\eqref{eq:Goyon}] and the other one to $f=0$. Moreover, it can be easily understood that, if the solid-like region is described by a small {\it yet non-vanishing} fluidity $f>0$, then it cannot remain solid forever and it will eventually flow, i.e., it is unstable. This situation therefore corresponds to transient shear banding as we shall describe in the following.

The third and last modeling step is to couple Eq.~\eqref{eq:fluidity_mobility} with an equation for the time evolution of the stress $\Sigma(t)$, which we suppose spatially homogeneous. This can be done by decomposing the total strain $\Gamma=\Gamma_{\mbox{\tiny el}}+\Gamma_{\mbox{\tiny pl}}$ into an elastic contribution $\Gamma_{\mbox{\tiny el}}= \tau \Sigma$, where  $\tau$ is a characteristic time inversely proportional to the elastic modulus, and a plastic contribution $\Gamma_{\mbox{\tiny pl}}$ such that $\dot{\Gamma}_{\mbox{\tiny pl}}=\Sigma \langle f \rangle$, where $\langle\dots\rangle$ denotes spatial average, and by using the well-known Maxwell model~\cite{Larson:1999}. The evolution equation for the shear stress then reads:
\begin{equation}\label{eq:Maxwell}
\frac{\dd \Sigma}{\dd t} = \frac{\dot\Gamma_{\mbox{\tiny el}}}{\tau} =  \frac{1}{\tau}\, [ \dot \Gamma - {\langle f \rangle} \Sigma]\,,
\end{equation}
with $\dot{\Gamma}=\dot{\Gamma}_{\mbox{\tiny el}}+\dot{\Gamma}_{\mbox{\tiny pl}}$.
This last equation is coupled to Eq.~\eqref{eq:fluidity_mobility}, which can be rewritten as: 
\begin{equation}\label{eq:fluidity} 
\frac{\partial f}{\partial \tilde t} =  f \left( \xi^2 \Delta f + m f - f^{3/2}\right)\,,
\end{equation}
with $m$ given by Eq.~\eqref{eq:m}.

Finally, we must specify boundary and initial conditions. Assuming that the external driving is acting at the boundary $y=0$, we choose the following boundary conditions. For $m^2>0$, we impose a ``wall fluidity'' $f_w=f(0,\tilde t) = m^2(\Sigma)$ at the moving wall, and $\partial_y f(L,\tilde t) = 0$ at the fixed wall. When $m=0$, we assume $\partial_y f(0,\tilde t) = 0 = \partial_y f(L,\tilde t)$ at both walls. Moreover, we take the initial fluidity profile to be homogeneous and very small, i.e., $f(y,0)=f_0$ with $f_0\ll 1$. As soon as $\Sigma >1$, we expect a shear band to develop from $y=0$ with a size $\lb(t)$ that increases with time and whose dynamics is set by the spatio-temporal evolution of the fluidity. 

Figure~\ref{fig:band_evolution} provides examples of numerical resolutions that illustrate the general phenomenology of shear start-up in our fluidity model. In particular, the time evolution of $\Sigma(t)$, obtained from the numerical integration of Eqs.~\eqref{eq:Maxwell} and~\eqref{eq:fluidity} for $\xi = 0.04$, $f_0 = 10^{-4}$, and $\dot \Gamma=2$, shows a stress overshoot very similar to that observed in experiments [Fig.~\ref{fig:band_evolution}(a)].
Moreover, the fluidity profiles $f(y, \tilde t)$ displayed in Fig.~\ref{fig:band_evolution}(b) for two specific times $\tilde t$
present a sharp interface between a fluidized shear band for $y<\lb$, where $f \sim m^2$, and a solid-like region for $y>\lb$, where $f(y, \tilde t) = f_0$. Note that the shear band grows in size because of the instability of the solid-like region, while retaining a sharp interface at $f\simeq f_0$. This results from our requirement that $\kappa[f]=f$. More precisely, it is possible to estimate analytically the steepness of the interface at $f \simeq f_0$, shown in dash-dotted lines in Fig.~\ref{fig:band_evolution}(b), as $\partial_y f\vert_{y=\lb} /m^2 = (m/5\xi^2)^{1/2}$. Deep into the region $f=f_0\ll 1$, i.e., far enough from the interface, the fluidity increases algebraically in time as $ f(y,\tilde t) \simeq  f_0 \left(1+f_0 \int_0^{\tilde t} m(s) \dd s \right)$. However, at the boundary of the fluid-like region for $y \simeq \lb(t)$, an instability occurs with an exponential growth of the fluidity. Dimensional considerations suggest that the instability extends over a scale of order $\xi/m^{1/2}$ and grows with a characteristic time scale $m^{-3}$. This implies that the size of the shear band $\lb(\tilde t)$ satisfies the equation~\cite{Benzi:2021PRE}:
\begin{equation}\label{eq:band_evolution}
\frac{\dd\lb}{\dd \tilde t} \sim \xi m^{5/2}.
\end{equation}
We will come back to the dynamics of the transient shear band below when discussing the fluidization. 

\section{Stress overshoot}

We now investigate how the stress maximum $\Sm$ scales with $\dot \Gamma$, which is an observable classically extracted from experiments~\cite{Carrier:2009,Divoux:2011,Fernandes:2017}. Concomitantly to the stress overshoot, the size $\lb$ of the shear band increases with time. As detailed in Ref.~\cite{Benzi:2021PRE}, at short time $\tilde t $, the band dynamics is dominated by the diffusion term
$f_w \xi^2 \sim m^2 \xi^2$
and $\lb$ grows as $(m^2\xi^2 \tilde t)^{1/2}$, while for large enough $\lb$, it follows Eq.~\eqref{eq:band_evolution}.
The overall process is illustrated in Fig.~\ref{fig:band_evolution}(c) by plotting the distance to the yield stress, $\Sigma-1$, as a function of the effective shear rate $L \dot \Gamma/\lb$, which corresponds to the average shear rate in the fluidized band. Such a representation of the flow dynamics clearly highlights the separation between two different dynamical regimes: a short-time ``unsteady'' regime that strongly depends on the applied shear rate $\dot\Gamma$ and where the data fall well below the equilibrium HB curve, and another ``quasi-steady'' regime at longer times, where all data nicely collapse on the HB curve, including during the
transient shear-banding regime.

\begin{figure}[t]
  \centering
  \includegraphics[width=0.85\columnwidth]{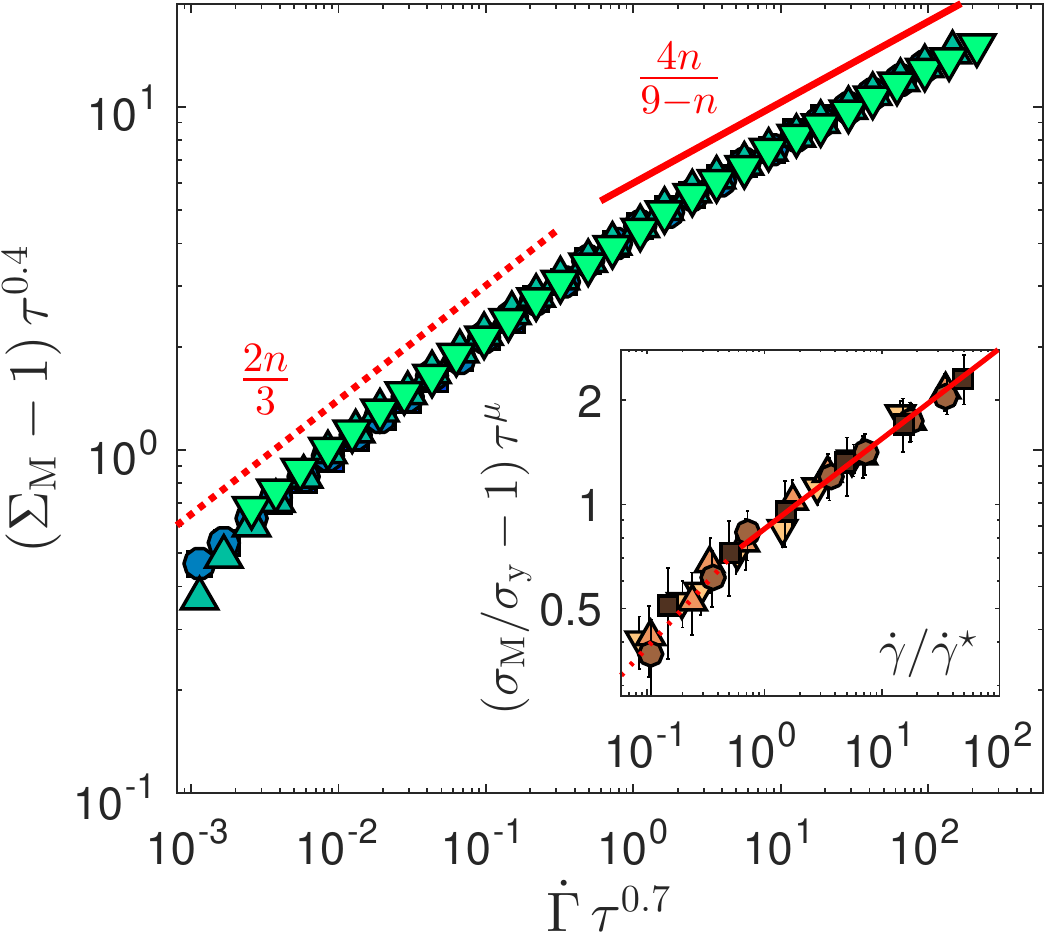}
  \caption{{\bf Scaling of the stress overshoot.} Predictions of the fluidity model for an HB exponent $n=1/2$. Rescaled stress maximum $(\Sm-1)\tau^{0.4}$ vs normalized shear rate $\Gp\tau^{0.7}$. Colored symbols refer to different values of $\tau$ from 0.1 to 100. Inset: Experimental results in Carbopol microgels. Rescaled stress maximum $(\sm/\sy-1)\tau^{\mu}$ vs normalized shear rate $\gp/\gp^\star$, where $\tau=\sy/G_0$ with $\sy$ and $G_0$ the yield stress and elastic modulus of the microgel respectively, $\mu=2n/(3-n)$, and $\gp^\star$ a rescaling factor as defined in Ref.~\cite{Benzi:2021PRL}. Colored symbols refer to different Carbopol concentrations from 0.1 to 3~\%~wt. The red solid lines (dotted lines resp.) show the scaling law inferred from the fluidity model with exponent $\alpha=4n/(9-n)$ in the asymptotic regime ($\beta=2n/3$ in the diffusive regime resp.).}
  \label{fig_overshoot_scaling}
\end{figure}

Based on the previous observations, the scaling of $\Sm$ with $\dot \Gamma$ can be computed using Eqs.~\eqref{eq:Maxwell} and~\eqref{eq:fluidity}. In a nutshell (see Refs.~\cite{Benzi:2021PRL,Benzi:2021PRE} for full details), considering that the stress grows linearly up to the stress maximum, i.e., $\Sigma(\tilde{t}) \simeq \tilde{t}/\tau$, that the stress is large enough that $m=(\Sigma-1)^{1/2n}/\Sigma^{1/2} \simeq \Sigma^{1/2n-1/2}$, and that $\langle f \rangle \sim \lb m^2$, the condition $\dd \Sigma /\dd\tilde t = 0$ at the stress maximum leads to:
\begin{equation}\label{eq:overshoot_master_eq}
\dot \Gamma = \left[\Sm(\titm)-1\right]^2 \lb(\titm)\,,
\end{equation}
where $\titm$ is the strain at the stress maximum $\Sm$. Further analysis of the two dynamical regimes then yields:
\begin{equation}\label{eq:Overshoot_Scaling}
\Sm-1 \sim B \left( \frac{\dot \Gamma}{\xi \tau^{1/2}} \right)^{\beta(n)} + C \left( \frac{\dot \Gamma}{\xi \tau} \right)^{\alpha(n)}\,, 
\end{equation}
where $B$ and $C$ are two numerical prefactors, and
\begin{equation}\label{eq:exponent} 
\beta(n) = \frac{2n}{3} \,\,\, \mathrm{and} \,\,\, \alpha(n) = \frac{4n}{9-n}\,.
\end{equation}
The first term on the r.h.s of Eq.~\eqref{eq:Overshoot_Scaling} dominates for small $\dot \Gamma$, when the shear band grows due to the diffusion term $f_w \xi^2$, while the second term dominates for large $\dot \Gamma$, when the shear band increases according to Eq.~\eqref{eq:band_evolution}. Note that Eqs.~\eqref{eq:Overshoot_Scaling} and~\eqref{eq:exponent} hold both for transient and stable shear bands. As shown in Ref.~\cite{Benzi:2021PRE}, an extensive survey of the existing numerical and experimental data shows excellent agreement with Eq.~\eqref{eq:Overshoot_Scaling}. Figure~\ref{fig_overshoot_scaling} illustrates this agreement by comparing the model predictions to experiments on Carbopol microgels. In both cases, two power-law regimes can be identified in $\Sm-1$ vs $\dot\Gamma$, and the exponents are consistent with the values $\beta=1/3$ in the ``diffusive'' regime and with $\alpha=4/17$ in the ``asymptotic'' regime at large $\dot\Gamma$ predicted for a shear-thinning index $n=1/2$. Finally, we emphasize that Eq.~\eqref{eq:Overshoot_Scaling} depends on $\xi$ with a singular limit for $\xi \rightarrow 0$, and that the above results depend on the choice $\kappa[f] \sim f$. The good agreement between Eq.~\eqref{eq:Overshoot_Scaling} and experimental data therefore provides strong support for the present formulation of the fluidity model, which constitutes a remarkable, non-trivial result.

\section{Including elasto-hydrodynamic (EHD) interactions into the model}
Based on experiments and numerical simulations, Cloitre, Bonnecaze and collaborators~\cite{Seth:2008,Seth:2011,Liu:2018} have shown that the flow of SGMs constituted of dense assemblies of deformable particles, such as microgels, emulsions or glasses of elastomeric particles, is controlled by {\it elastohydrodynamic} (EHD) interactions, which result from the lubrication flows of solvent within the thin films between the particles.  
In particular, a recent study~\cite{Khabaz:2021} has shown that EHD interactions impact the scaling of the stress overshoot in a non-trivial way. We herewith discuss an easy way to include such EHD effects in our continuum model through a simple modification of the Maxwell equation \eqref{eq:Maxwell} for the stress evolution. We propose to add a contribution $\Gamma_{\mbox{\tiny EHD}}$ from EHD interactions to the total strain,  $\Gamma=\Gamma_{\mbox{\tiny el}}+\Gamma_{\mbox{\tiny pl}}+\Gamma_{\mbox{\tiny EHD}}$, which is related to the shear stress through 
$ \dot{\Gamma}_{\mbox{\tiny EHD}}=\dot{\Gamma}_0 \Sigma^{2}$, 
where $\dot{\Gamma}_0$ is a reference shear rate below which EHD effects become significant. This specific choice of scaling for the EHD interactions is justified by~\cite{Meeker:2004a,Seth:2011}. The resulting modified Maxwell model reads:
\begin{equation}\label{eq:Maxwell_new}
\frac{\dd \Sigma}{\dd \tilde t} = \frac{1}{\tau}\left[ 1 - \frac{{\langle f \rangle} \Sigma}{\dot \Gamma}-\dot{\Gamma}_0\frac{\Sigma^2}{\dot{\Gamma}}\right].
\end{equation}

\begin{figure}[t]
  \centering
  \includegraphics[width=1\columnwidth]{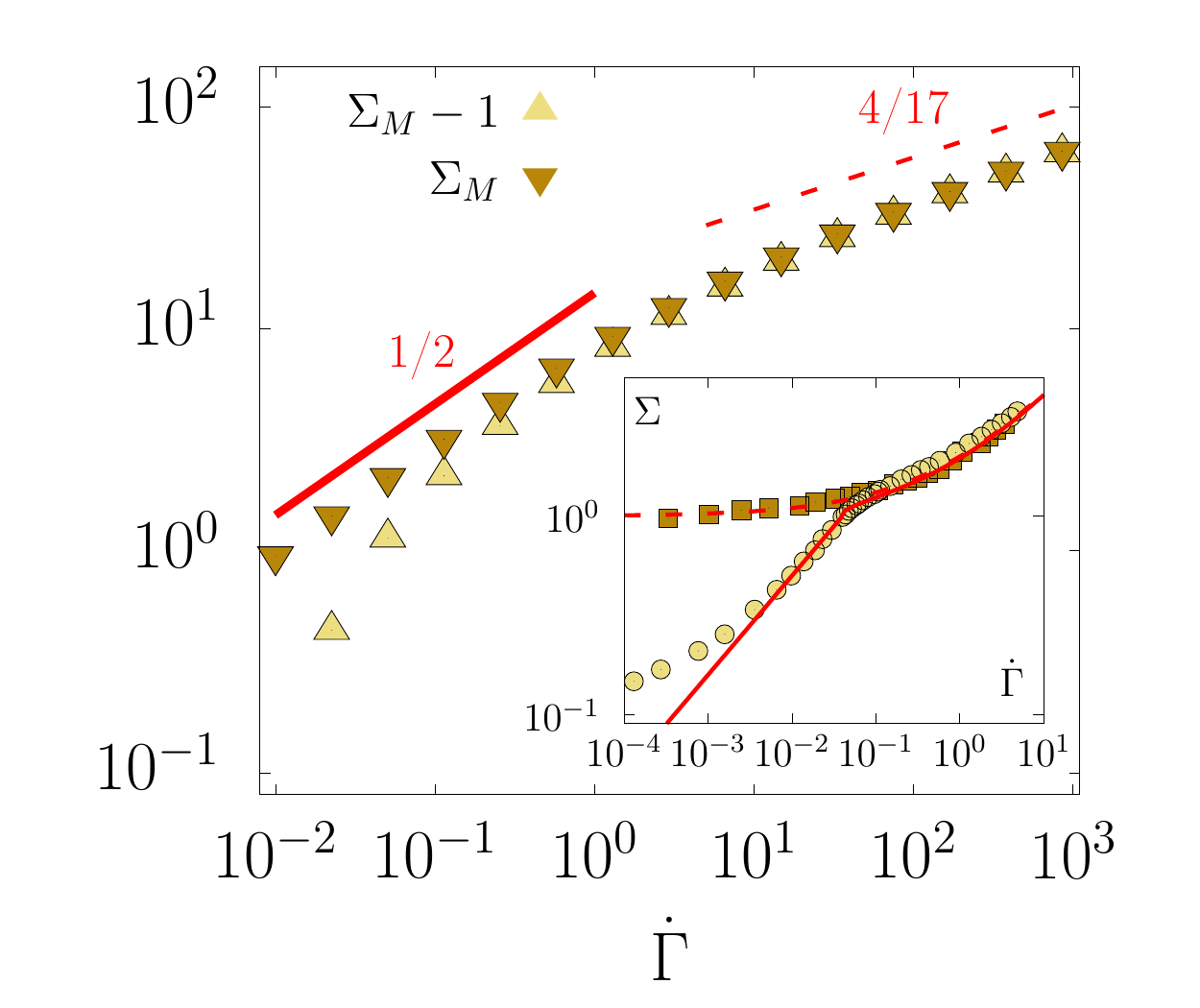}
  \caption{{\bf Effects of elasto-hydrodynamic (EHD) interactions}. 
   Scaling of the stress overshoot maximum $\Sm$ (\protect \invertedtriangledarkgoldenrod) and its distance to the yield stress $\Sm-1$ (\protect \trianglelightgoldenrod) as a function of the shear rate $\dot{\Gamma}$ computed from the fluidity model with  $n=1/2$, $\dot{\Gamma}_0=0.01$, and $\xi=0.001$. The red solid and dashed lines show the scaling laws predicted by the model respectively with EHD interactions (exponent 1/2) and without EHD interactions (exponent 4/17). Insert: Steady-state flow curve for microgels on smooth (\protect \circlelightgoldenrod) and rough (\protect \squaredarkgoldenrod) surfaces extracted from Fig.~1 of Ref.~\cite{Meeker:2004a}. The red dashed line is the HB flow curve with $n=1/2$, i.e., $\Sigma=1+\dot{\Gamma}^{1/2}$, while the red solid line is the flow curve computed from Eq.~\eqref{eq:EHD_stationary} with $\dot{\Gamma}_0=0.04$.
  }
  \label{fig:HB_EHD}
\end{figure}

First, EHD interactions modify the steady-state rheology. Indeed, with $f=m^2\,\Theta (\Sigma-1)$ and $\dd \Sigma/\dd \tilde t=0$, we get:
\begin{equation}\label{eq:EHD_stationary}
\dot{\Gamma}=(\Sigma-1)^{1/n} \Theta(\Sigma-1)+\dot{\Gamma}_0 \Sigma^2\,.
\end{equation}
The inset in Fig.~\ref{fig:HB_EHD} compares the steady-state flow curve predicted from Eq.~\eqref{eq:EHD_stationary} with $n=1/2$ and $\dot{\Gamma}_0=0.04$ to experimental data on microgels obtained under two different boundary conditions~\cite{Meeker:2004a}. 
While the experimental flow curve for rough shearing surfaces (brown squares) nicely follows the HB law, the flow curve measured for smooth surfaces (yellow circles) presents a kink for $\dot{\Gamma}\lesssim 0.05$ that is usually interpreted as the hallmark of predominant slippage at the walls~\cite{Bonn:2017}. Interestingly, including EHD interactions in our model using $\dot{\Gamma}_0=0.04$ allows us to nicely predict the steady-state flow curve for the smooth surface: in spite of some deviations at extremely low shear rates, EHD contributions produce deviations from the HB behaviour when $\dot{\Gamma} \lesssim \dot{\Gamma}_0$, while leaving the HB flow curve essentially unaltered for $\Sigma \gg 1$, very much like experimental observations. This suggests that $\dot{\Gamma}_0$ most probably embeds some non-trivial dependence on boundary roughness, which remains to be modeled theoretically.

Second, EHD interactions also modify the scaling of the stress overshoot, which can be derived using Eqs.~\eqref{eq:fluidity} and~\eqref{eq:Maxwell_new}. In particular, EHD interactions change Eq.~\eqref{eq:overshoot_master_eq} to:
\begin{equation}\label{eq:overshoot_master_eq_new}
\dot \Gamma = \left[\Sm(\titm)-1\right]^2 \lb(\titm)+\dot{\Gamma}_0\Sm^2\,.
\end{equation}
Therefore, for $\dot\Gamma<\dot\Gamma_0$ where EHD effects dominate, 
the stress maximum no longer depends on the HB exponent, but rather simply on the EHD scaling as $\Sm \sim ({\dot{\Gamma}}/{\dot{\Gamma}_0})^{1/2}$.
For $\dot{\Gamma} \gg \dot{\Gamma}_0$, however, the scaling of Eqs.~\eqref{eq:Overshoot_Scaling} and~\eqref{eq:exponent} is recovered. This is confirmed in Fig.~\ref{fig:HB_EHD} by the numerical integration of the full dynamical equations. Note that when EHD interactions dominate, the  exponent $1/2$ is observed for the stress maximum $\Sm$ rather than for $\Sm-1$, which indicates that the yield stress is no longer a ``reference'' stress for the stress overshoot. 

\section{Transient shear banding and fluidization time}
The fluidity model can be further used to compute the duration of the transient shear banding regime, i.e., the fluidization time $T_\text{f}$ as a function of $\dot \Gamma$, which constitutes an important prediction for experiments and applications of SGMs. Indeed, during the fluidization process, the system satisfies the balance $\dot \Gamma = \langle f \rangle \Sigma = m^2 \lb \Sigma$. This allows us to compute $m$ as a function of $\lb$ and $\dot \Gamma$. Using Eq.~\eqref{eq:band_evolution}, one then predicts that, for small enough $\dot \Gamma$:
\begin{equation}\label{eq:Tfscaling}
T_\text{f} \sim \frac{1}{\xi \dot \Gamma^{9/4}}\,,
\end{equation}
in excellent agreement with experimental data
\cite{Divoux:2010,Benzi:2019}. Note that the scaling exponent for $T_\text{f}$ vs $\dot \Gamma$ is independent of the HB exponent $n$. 


In the case of stress-induced fluidization, i.e., when forcing at a constant external stress $\Sigma$, Eq.~\eqref{eq:band_evolution} can still be used upon identifying $\tilde t = m^2 t$, which results from the fact that $m^2 \sim \dot \Gamma$ for small $\dot \Gamma$. Thus, Eq.~\eqref{eq:band_evolution} generally predicts $T_\text{f} \sim (\xi m^{9/2})^{-1}$, 
which leads to $T_\text{f} \sim 1/[\xi (\Sigma-1)^{9/4n}]$ 
for small imposed values of $\Sigma-1$. Therefore, the present fluidity model predicts that the ratio of the scaling exponents under imposed $\dot\Gamma$ to that under imposed $\Sigma$ is given by the HB exponent $n$, as observed in experiments on Carbopol microgels \cite{Divoux:2011b,Benzi:2019}. 

Finally, as examined in details in Ref.~\cite{Benzi:2021PRE}, one may introduce long-range correlations in the fluidity through noise-like dynamics and investigate how the above predictions depend on boundary conditions. In brief, the transient shear-banding scenario and the scaling of $T_\text{f}$ given by Eq.~\eqref{eq:Tfscaling} are very robust to fluidity correlations when the fluidity at the moving wall is fixed through $f_w=f(0,\tilde{t})=m^2$. However, when rather fixing the fluidity gradient at the moving wall through $\partial_{y} f(0,0)=0$, long-range spatial correlations conspire with the boundary condition to promote the emergence of a completely different fluidization scenario, where the growth of the shear band is prevented, leading to a stress increase that is initially smoother, but later characterized by an abrupt drop, resembling brittle-like failure~\cite{Magnin:1990} and similar to the one discussed in recent theoretical and numerical works~\cite{Ozawa:2018,Singh:2020,Barlow:2020}.

\section{Summary and open questions}
We started this Perspective paper by asking how two classical observables that characterize shear start-up in SGMs, namely the stress overshoot $\Sm=\sm/\sy$ and the fluidization time, depend upon the applied shear rate $\dot\Gamma$. We have shown that a dynamical fluidity model allows one to predict the way rheological variables should be analyzed, i.e., $\Sm-1$ vs. $\dot{\Gamma}$ and $\Tf$ vs. $\dot{\Gamma}$ or $m$. The corresponding scaling exponents and their dependence on the HB exponent are in excellent agreement with experiments on Carbopol microgels. The model is versatile enough to include EHD interactions that, when dominant, change the scaling of the stress overshoot to $\Sm\sim\dot\Gamma_0$ whatever the underlying HB behaviour. Overall, the model predictions hinge on some basic ingredients: (i)~the ``mobility'' function $\kappa[f]$ in Eq.~\eqref{eq:fluidity_mobility} that allows the coexistence of a fluidized band and a solid-like region, (ii)~the Maxwell equation \eqref{eq:Maxwell} for the stress evolution, and (iii)~the boundary conditions, which are crucial, for they discriminate between ductile-like and brittle-like types of fluidization. Focusing on items (ii) and (iii) above, we highlight two important open problems. 

First, future work should analyze situations where the forcing has some increased complexity. In particular, it would be very interesting to explore the present fluidity model with {\it time-dependent} protocols such as the shear-rate ramps that are widely used by rheologists. Whether or not this model may predict rheological hysteresis in SGMs and its dependence with the shear-rate sweep rate \cite{Divoux:2013,Radhakrishnan:2017,Jamali:2019}, in the two cases of transient and permanent shear banding, is an outstanding task.

Second, accounting precisely for boundary conditions is key for further theoretical advances. As already noted in Ref.~\cite{Benzi:2021PRE}, since a simple change of boundary conditions may suppress the nucleation of the fluid-like phase at the moving wall, boundary conditions appear to control the shear-induced solid-to-liquid transition in SGMs. However, we still miss physical insight into the microscopic dynamical processes at play at the walls. Here, the proposed phenomenological treatment of EHD interactions and the observation that EHD parameters must depend on boundary conditions call for more modelling effort.
This can open the way to obtain a realistic fit of experimental results once the various parameters for the continuum modelling are extracted from experiments.

\begin{acknowledgments}
The authors thank Thomas van Vuren for insightful discussions. 
\end{acknowledgments}


\section*{References}

%

\end{document}